# On Search for New Physics in Nonequilibrium Reactor Antineutrino Energy Spectrum

V. I. Kopeikin

*Russian Research Centre Kurchatov Institute, pl. Kurchatova 1, Moscow, 123182 Russia*

**Abstract**—The calculations of the time-dependent reactor antineutrino energy spectrum are presented. Some problems associated with sensitive searches for neutrino magnetic moment and neutrino oscillations in reactor antineutrino flux are considered.

## 1. INTRODUCTION

The previous results from $\bar{\nu}_e e$ scattering reactor experiments were interpreted as an upper limit of $2\times10^{-10}\,\mu_B$ ($\mu_B$ is electron Bohr magneton) on the neutrino magnetic moment $\mu_\nu$ [1]. Direct measurements of the neutrino magnetic moment at level $\sim 10^{-11}\,\mu_B$ would have a serious impact on particle physics and astrophysics [2]. To achieve this sensitivity the low energy recoil electrons in $\bar{\nu}_e e$ scattering should be detected. Preparations for such experiments are under way [3]. As was discussed in [4], the soft part of the recoil electron energy spectrum may be strongly time-dependent during reactor cycle. In the present study the time evolution of the $\bar{\nu}_e e$ scattering cross sections expected in reactor experiments for the weak and magnetic interactions are calculated.

In any neutrino reaction, the cross section observed $\sigma_{fis}$ is the reaction cross section for monoenergetic antineutrinos $\sigma(E_\nu)$ (cm$^2/\bar{\nu}_e$) folded with reactor $\bar{\nu}_e$ spectrum $\rho(E_\nu,t)$ ($\bar{\nu}_e$/MeV fission)

$$\sigma_{fis}(t)=\int\rho(E_\nu,t)\,\sigma(E_\nu)\,dE,\ [\text{cm}^2\text{fission}^{-1}]. \quad (1)$$

To interpret data, we, therefore, must have an accurate knowledge of the antineutrino spectrum. In this work the typical thermal power reactor $\bar{\nu}_e$ spectrum and its time evolution are calculated.

Oscillation experiments are based on the reaction

$$\bar{\nu}_e + p \rightarrow e^+ + n. \quad (2)$$

Any distortions of the positron energy spectrum or decrease of the cross section (1), measured at reactor would indicate oscillations. In a recent paper [5] we calculated corrections to the non-oscillation cross section for reaction (2) that had been precisely measured near power reactor [6]. In a present study we discuss the role of the residual $\bar{\nu}_e$ emission after reactor shutdown in measuring of the positron spectrum.

## 2. REACTOR ANTINEUTRINO SPECTRUM

Here we consider the antineutrino energy spectrum of a Light Water Reactor (LWR). Reactors of this type are used in the most neutrino experiments. They usually operate for 11 months, followed by a shutdown of one month for one-third of fuel elements replacement. At the beginning of each annual reactor cycle $\alpha_5=69\%$ of the fissions are from $^{235}$U, $\alpha_9=21\%$ from $^{239}$Pu, $\alpha_8=7\%$ from $^{238}$U, and $\alpha_1=3\%$ from $^{241}$Pu. During operation $^{235}$U burn up and $^{239}$Pu and $^{241}$Pu are accumulated from $^{238}$U. The average ("standard") fuel composition is

$$\bar{\alpha}_5=58\%,\ \bar{\alpha}_9=30\%,\ \bar{\alpha}_8=7\%,\ \bar{\alpha}_1=5\%. \quad (3)$$

The present calculation of the time-dependent reactor $\bar{\nu}_e$ spectrum during operation and shutdown periods include:

a. Time-dependent $\bar{\nu}_e$ activity of the fission products, including possible isomeric states, effect of delayed neutrons and transmutations of fission products by reactor neutrons. It has been taken into account the time-dependent fission contributions $\alpha_i(t)$ (i=5,9,8,1) both current and two previous annual reactor cycles.

b. Time-dependent $\bar{\nu}_e$ activity coming from the neutron captures by heavy elements. A dominant contribution comes from the captures by $^{238}$U:
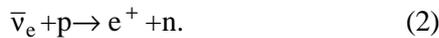
$^{238}$U(n,$\gamma$)$^{239}$U$\rightarrow$$^{239}$Np$\rightarrow$$^{239}$Pu.

Relative contribution of the other antineutrino sources in the reactor core is less than 0.5% [7].

The calculation of the $\bar{\nu}_e$ spectrum below 2 MeV (to which three-fourths of all emitted antineutrinos belong) and its components is presented in Fig.1. The $\bar{\nu}_e$ spectrum in the energy range of $E_\nu=2$ - 9 MeV has been measured at Rovno reactor [7]:

$$\rho(E_\nu)=5.09\cdot\exp[-(E_\nu/1.54)-(E_\nu/6.05)^2-(E_\nu/7.73)^{10}],\ [\text{MeV fission}]^{-1}. \quad (4)$$

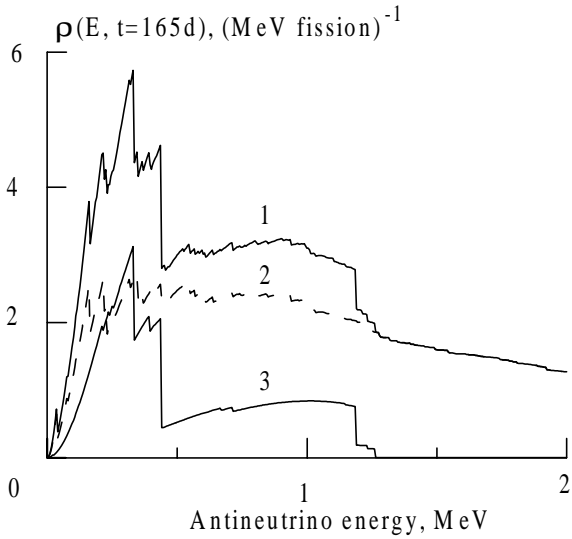

Fig. 1. Energy spectrum of antineutrinos from LWR reactor at the middle of the 330-d operating period: 1 – all antineutrinos, 2 – fission antineutrinos, 3 – antineutrinos, associated with neutron captures in heavy elements.

The spectra, both calculated (Fig.1) and measured (4) corresponds to standard fuel composition (3). The typical time evolution of the antineutrino spectrum during annual reactor cycle is shown in Fig. 2 and Fig. 3.

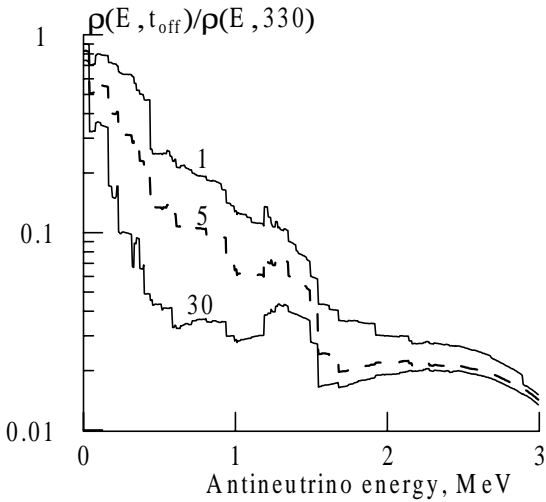

Fig.2. Ratios of the current $\bar{\nu}_e$ spectra after reactor is shut down to the spectrum at the end of the 330-d reactor operation period. The numbers on the curves indicate days after reactor shutdown.

This evolution during reactor operation is mainly caused by accumulation of the nuclei beta activity (in the region E< 2 MeV) and changes in the reactor fuel composition (in the region E> 2 MeV).

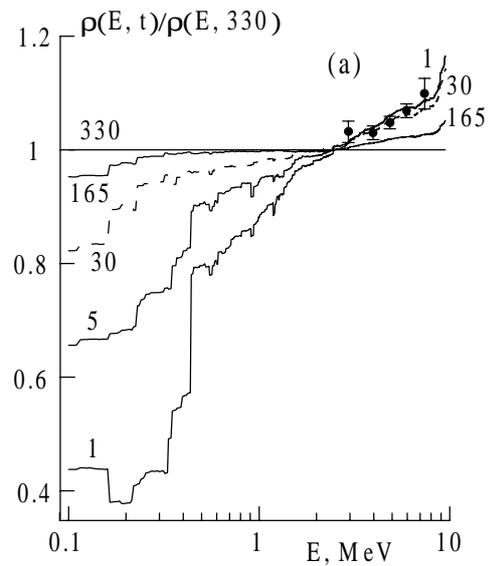

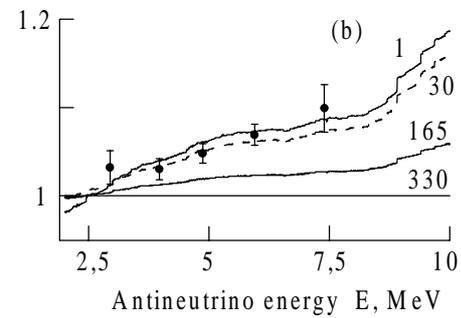

Fig. 3. Ratios of the current reactor $\bar{\nu}_e$ spectra during reactor operation to that at the end of the 330-d reactor operating period: lines – present calculation, circles – experiment at Rovno reactor [7]. The numbers at the curves indicate days from the beginning of the operating period.

## 3. CROSS SECTIONS FOR SCATTERING OF REACTOR ANTINEUTRINOS ON ELECTRONS

The calculated weak and magnetic elastic cross sections $\sigma_{fis}$ (1) for the $\bar{\nu}_e e$ scattering reaction as a function of recoil electron energy are shown in Fig. 4. Corrections associated with the electron binding in atoms were considered in [8]. It should be emphasized that, weak $\bar{\nu}_e e$ scattering plays the role of a background, which must be exactly calculated and subtracted. In sensitive searches for $\mu_\nu$, the measured weak recoil electron energy spectrum could be used as a tool for detector check and calibration. The calculated time variations of the folded cross sections (1) for weak and magnetic $\bar{\nu}_e e$ scattering during reactor cycle are presented in Fig. 5.

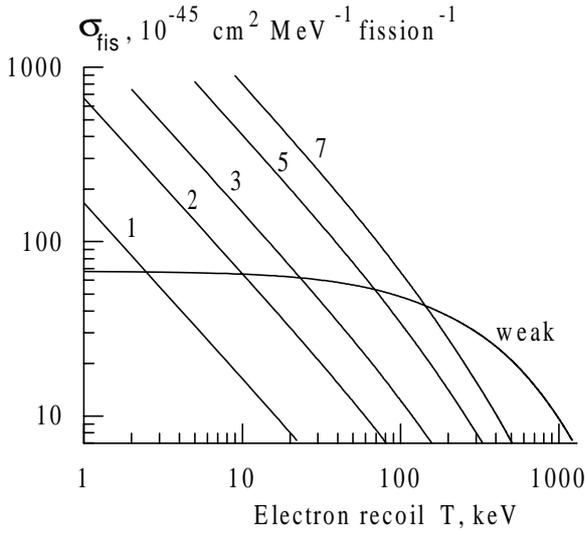

Fig. 4. Cross sections for weak and magnetic scattering of reactor antineutrinos on free electrons at the middle of the 330-d operating period. The numbers on the curves indicate the values of the moment $\mu_\nu$ in $10^{-11} \mu_B$.

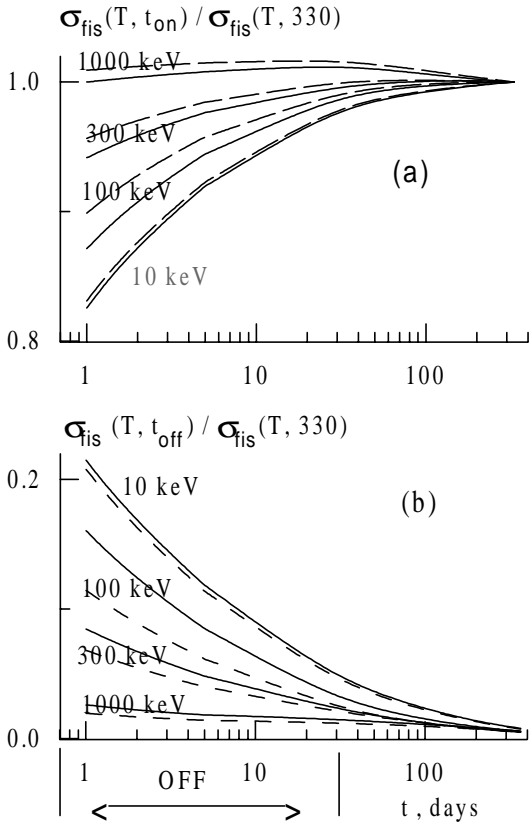

Fig. 5. Ratios of the current $\bar{\nu}_e e$ scattering cross sections expected in experiment to those at the end of the 330-d reactor operation period for the reactor (a) operating and (b) shutdown (OFF) periods. Data are presented for the four groups of recoil electron energies. The solid (dashed) lines represent magnetic (weak) scattering.

## 4. ABOUT PRECISION MEASUREMENT OF POSITRON SPECTRUM FOR THE REACTON

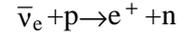

$$\bar{\nu}_e + p \to e^+ + n$$

Radical improvements of the detector characteristics including essential decrease of the accidental and correlated backgrounds have been achieved in the last long baseline reactor oscillation experiments. In the present study we discuss the third type of background which did not take earlier into account in measuring of positron spectrum. It is a reactor correlated positron background, associated with residual $\bar{\nu}_e$ emission after reactor shutdown. This positron background is about 3% in the energy range $T_{e^+} < 1.2$ MeV, as regards to positron rate when reactor in operation, see Fig. 6.

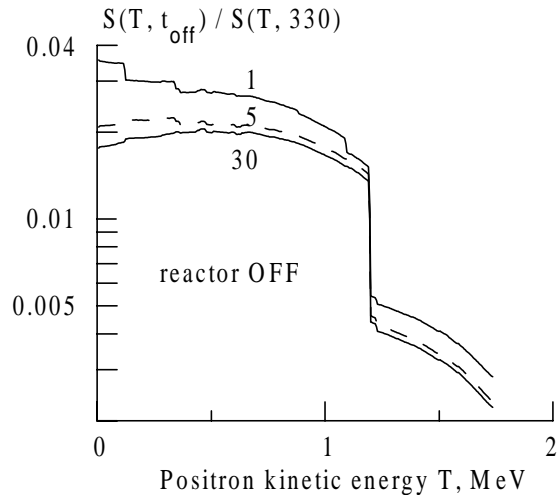

Fig. 6. Ratios of the current positron spectra for the reaction $\bar{\nu}_e + p \to e^+ + n$, associated with residual antineutrino emission after reactor is shut down, to the spectrum at the end of 330-d reactor operation period. The numbers at the curves indicate days after reactor shutdown.

This effect may be more significant if oscillation experiment is implemented by one detector positioned from near (distance r) and fare (distance R) reactors. Such an experimental setting was implemented, for example, in Bugey, r=15 m and R=95 m [9]. In this situation the positron background signal from the near stopped reactor $S_{near}^{OFF}$ for $T_{e^+} < 1.2$ MeV is approximately equal positron signal from the operating fare reactor $S_{far}^{ON}$, that is:

$S_{near}^{OFF} / S_{far}^{ON} \sim 1$ for $T_{e^+} = 0 - 1.2$ MeV and

$\sim 0.15$ for $T_{e^+} = 1.2 - 1.7$ MeV.

## 5. CONCLUSION

Searches for new physics in neutrino experiments at nuclear reactors require refining our knowledge of the reactor $\bar{\nu}_e$ spectrum. At the present level of experimental accuracy and sensitivity a detailed and profound analysis of reactor $\bar{\nu}_e$ spectrum for each particular experiment should be carried out.

## ACKNOWLEDGMENTS

The author would like to thank L. A. Mikaelyan and V. V. Sinev for helpful discussions. This work was supported by the Russian Foundation for Basic Research (project nos. 00-15-06708 and 00-02-16035.)

## REFERENCES

1. A. V. Derbin, Yad. Fiz. **57**, 236 (1994) [Phys. At. Nucl. **57**, 222 (1994)].
2. L. B. Okun, M. B. Voloshin, M. I. Vysotsky, Zh. Eksp. Teor. Fiz. **91**, 754 (1986) [Sov. Phys. JETP **64**, 446 (1986)]; P. Vogel, J. Engel, Phys. Rev. D **39**, 3378 (1989).
3. L. A. Mikaelyan, Yad. Fiz. **65**, (2002), submitted.
4. V. I. Kopeikin, L. A. Mikaelyan, V. V. Sinev, Yad. Fiz. **63**, 1087 (2000) [Phys. At. Nucl. **63**, 1012 (2000)].
5. V. I. Kopeikin, L. A. Mikaelyan, V. V. Sinev, Yad. Fiz. **64**, 914 (2001) [Phys. At. Nucl. **64**, 849 (2001)].
6. V. N. Vyrodov, Y. Declais, H. de Kerret et. al., Pis'ma Zh. Eksp. Teor. Fiz. **61**, 161 (1995) [JETP Lett. **61**, 163 (1995)].
7. V. I. Kopeikin, L. A. Mikaelyan, V. V. Sinev, Yad. Fiz. **60**, 230 (1997) [Phys. At. Nucl. **60**, 172 (1997)].
8. V. I. Kopeikin, L. A. Mikaelyan, V. V. Sinev and S. A. Fayans, Yad. Fiz. **60**, 2032 (1997) [Phys. At. Nucl. **60**, 1859 (1997)].
9. B. Achkar, R. Aleksan, M. Avenier et.al. Nucl. Phys. B **434**, 503 (1995).